\documentclass[a4paper,11pt]{article}
\pdfoutput=1

\usepackage[utf8]{inputenc}
\usepackage[T1]{fontenc}
\usepackage[english]{babel}
\usepackage{layout}
\usepackage{setspace}
\usepackage{lmodern}
\usepackage{enumitem}
\usepackage{soul}
\usepackage{eurosym}

\usepackage{verbatim}
\usepackage{moreverb}

\usepackage{sectsty}

\usepackage{mathrsfs}
\usepackage{cancel}

\usepackage{makeidx}
\usepackage{multirow}
\usepackage{stmaryrd}

\usepackage{docstyle}

\title{\textcolor{mycolor}{Extra Dimensions \& Fuzzy Branes\\in String-Inspired Nonlocal Field Theory}}

\author{Florian Nortier}

\affiliation{Université Paris-Saclay, CNRS, CEA, Institut de Physique Théorique,}
\affiliation{91191, Gif-sur-Yvette, France}

\emailAdd{florian.nortier@ipht.fr}

\abstract{Particle physics models with extra dimensions of space (EDS's) and branes shed new light on electroweak and flavor hierarchies with a rich TeV scale phenomenology. This article highlights new model building issues with EDS's and branes, arising in the framework of weakly nonlocal field theories. It is shown that a brane-localized field is still delocalized in the bulk on a small distance from the brane position: fields localized on such distant fuzzy branes are thus allowed to interact directly with suppressed couplings. Directions for model building are also given: (i) with fuzzy branes, a new realization of split fermions in an EDS is presented, naturally generating flavor hierarchies; (ii) with a warped EDS, the usual warp transmutation of a brane-localized mass term is revisited, where it is shown that the nonlocal scale is also redshifted and provides a smooth UV cutoff for the Higgs boson mass. This framework is expected to have natural UV completions in string theory, but the possibility to embed it in recent UV complete weakly nonlocal quantum field theories is commented.}

\keywords{Braneworlds, Weak Nonlocality, String Phenomenology, Hierarchy Problem, Flavor Models}

\arxivnumber{2112.15592}

\begin{document}

\maketitle
\flushbottom


\section{Introduction}
\label{introduction}
Despite its great experimental success \cite{Workman:2022ynf}, the SM should not be the end of the high-energy physics (HEP) story, since we have several good reasons to go beyond the SM (BSM) \cite{Nagashima:2014tva}. From a bottom-up perspective, HEP models based on local quantum field theories (QFT's) with extra dimensions of space (EDS's) and branes were extensively used to solve or reformulate various BSM issues, with a very rich phenomenology \cite{Raychaudhuri:2016}. The price to pay for models where some SM fields propagate in EDSs is that gauge theories are not perturbatively renormalizable anymore. From a top-down perspective, these constructions are usually considered as effective field theories (EFT's) that may have UV completions in string theory \cite{Becker:2007zj, Ibanez:2012zz}.\\

Since strings are 1D extended objects, string theory leads naturally to nonlocal effects at the string scale $M_s$, such that these effects decrease rapidly for distances larger than the string length $\ell_s = 1/M_s$ \cite{Marshakov:2002ff}. In the perturbative worldsheet formulation, the modular invariance is at the origin of a UV/IR mixing induced by nonlocality \cite{Abel:2021tyt, Abel:2023hkk}. It reduces to a worldline inversion symmetry in the particle limit, and one gets infinite-derivative operators in the effective Lagrangian that map the UV divergences to IR ones \cite{Abel:2019ufz, Abel:2019zou, Abel:2020gdi}. Moreover, $T$-duality implies nonlocality \cite{Padmanabhan:1996ap}, and little string theories are known to be quasilocal field theories \cite{Kapustin:1999ci}. 
In the nonperturbative formulation offered by string field theory (SFT) \cite{Erbin:2021smf}, nonlocality is already manifest in the Lagrangian via infinite-derivative operators \cite{Ohmori:2001am, Arefeva:2001ps}. Analyticity and unitarity follows from Efimov analytic continuation \cite{Efimov:1965mnl, Efimov:1966ylf, Pius:2016jsl, DeLacroix:2018arq}.
During these recent years, such stringy nonlocal effects have been implemented in 4D bottom-up models via exponential form factors \cite{Buoninfante:2018mre}, in order to give some insights on how string theory softens the UV behavior in particle physics \cite{Biswas:2014yia, Ghoshal:2017egr, Ghoshal:2020lfd, Frasca:2020jbe, Frasca:2020ojd, Frasca:2021iip, Mo:2022szw, Chatterjee:2023ehr}. The reader could refer to Refs.~\cite{Buoninfante:2022ild, Koshelev:2023elc} for reviews on other applications in black hole physics and cosmology.\\

The purpose of this article is to investigate for the first time the interests in such weakly nonlocal (WNL) effects in models with EDS's and branes. In a local higher-dimensional EFT, thin\footnote{A thin or $\delta$-like brane is a brane with zero thickness (in the regime of validity of the EFT).} (or $\delta$-like) 3-branes are perfectly acceptable objects in the worldvolume of which 4D fields are trapped \cite{Sundrum:1998sj}. However, locality requires that the brane-localized interaction/kinetic terms appear as pointlike in the transverse EDS's, which sometimes creates ill-defined situations \cite{Goldberger:2001tn, Lewandowski:2001qp, delAguila:2003bh, Kyae:2003nc, delAguila:2004xd, Olechowski:2008bh, Carena:2012fk, Fichet:2019owx}, and requires a careful mathematical treatment \cite{delAguila:2006atw, deRham:2007mcp, Angelescu:2019viv, Nortier:2020lbs, Nortier:2020xms, Nortier:2020vge, Leng:2020ofk}. In a WNL framework, it is clear that point interactions on branes are smeared by the delocalized vertices on $\delta$-like branes. For instance, it has been shown that WNL resolves the transverse singular behavior of $\delta$-like sources \cite{Boos:2018kir, Giacchini:2018wlf, Boos:2020ccj} like branes \cite{Boos:2018bxf, Boos:2020kgj}.\\

This article proposes a set of new interesting possibilities relying on WNL field theory that provide new paths to solve BSM issues in HEP models with EDS's and branes. The analysis is restricted to toy models to focus on the new features of weak nonlocality (WNL) at tree level: there is no claim to have achieved realistic BSM scenarios. The goal is to highlight new possible model building issues that could be used in realistic model building, instead of discussing the detailed phenomenology of specific models. For that purpose, it is sufficient to limit the study to the example of SFT-like form factors, motivated by a bottom-up approach towards UV completions in string theory, but the developed methods can be extended to other WNL form factors from some other nonlocal UV theories.\\

The article is organized as follows.
In Section~\ref{nonlocal_braneworld}, a WNL toy model with bulk and brane-localized scalar fields is studied, and the inclusion of gauge symmetries is discussed. The goal is to provide, for the first time, tools for building WNL braneworlds. Section~\ref{applications} provides 2 possible applications of WNL for bottom-up model building: (i) a split fermion scenario \cite{ArkaniHamed:1999dc, Mirabelli:1999ks} with multiple fuzzy branes; (ii) a warped extra dimension \cite{Randall:1999ee} with a redshifted WNL scale. The results are summarized in Section~\ref{conclusion}. In Appendix~\ref{conventions}, the notations and conventions used in all sections are specified.

\section{String-Inspired Nonlocal Braneworlds}
\label{nonlocal_braneworld}

\subsection{Weakly Nonlocal Scalar Fields}
In SFT, if one truncates the tower of string excitations to its lowest level, one is left with an EFT which has weakly nonlocal (WNL) form factors, e.g. to study tachyon condensation in open bosonic SFT \cite{Ohmori:2001am, Arefeva:2001ps}. The propagator has a single pole corresponding to a standard degree of freedom: such infinite-derivative field theory is ghost-free by construction \cite{Buoninfante:2018mre}. The WNL scale $\Lambda_\eta = 1/\eta$ acts as a cutoff in momentum space, where the stringy behavior of the particles starts softening the UV behavior of the dynamics $(\Lambda_\eta \sim M_s)$. Actually, these peculiar toy models are UV finite in perturbative QFT computations \cite{Tomboulis:2015gfa}, and the nonlocal scale is stable under radiative corrections. They are thus reminiscent of the old WNL QFT's to handle UV divergences \cite{Namsrai:1986md}. In this section, such effects 	are implemented in a bottom-up braneworld model.

\subsubsection{Toy Model}
\label{toy_model}

\paragraph{Fields \& Symmetries:}
Consider a 5D Euclidean (orbifolded) spacetime $\mathbb{E}^5 = \mathbb{R}^4 \times S^1 / \mathbb{Z}_2$, where $S^1$ is the circle of radius $\rho$. The field and symmetry content of the WNL toy model are given in the following list:
\begin{itemize}[label=$\spadesuit$]
\item 1 real 5D scalar field $\Phi(x, y)$ (mass dimension $3/2$) propagates into the whole bulk, and is even under the $\mathbb{Z}_2$ orbifold symmetry.
\item 2 real 4D scalar fields $\theta(x)$ and $\omega(x)$ (mass dimension 1) are localized respectively on the branes at $y = 0, \pi \rho$. From these 4D fields, one defines the localized ``5D fields'' $\Theta(x, y) = \theta(x) \, \delta(y)$ and $\Omega(x, y) = \omega(x) \, \delta(y - \pi \rho)$, which describe 4D degrees of freedom sharply localized on the 3-branes.
\item The model has an exchange symmetry between the brane fields $\theta(x) \leftrightarrow \omega(x)$.
\end{itemize}

\paragraph{String-Inspired Nonlocality:}
Following the standard construction of string-inspired nonlocal field theories \cite{Buoninfante:2018mre}, the following 5D smeared fields $\widetilde{S} = \widetilde{\Phi}, \widetilde{\Theta}, \widetilde{\Omega}$ are defined by acting a SFT-like smearing operator on the local 5D fields $S = \Phi, \Theta, \Omega$, such that
\begin{equation}
\widetilde{S}(x, y) = e^{\eta^2 \Delta} \, S(x, y) \, .
\end{equation}
There is a UV scale $\Lambda_{UV}$, which is identified with the usual UV cutoff of a braneworld EFT in the local limit $\eta \rightarrow 0$.
In the case of the quasilocalized 5D fields, one can also split them as $\widetilde{\Theta}(x, y) = \widetilde{\theta}(x) \, \delta_\eta(y)$ and $\widetilde{\Omega}(x, y) = \widetilde{\omega}(x) \, \delta_\eta(y - \pi \rho)$, with
\begin{equation}
\widetilde{s}(x) = e^{\eta^2 \Delta_\parallel} \, s(x) \, , \ \ \ \delta_\eta(y) = e^{\eta^2 \Delta_\perp} \, \delta(y) \, ,
\end{equation}
where $s = \theta/\omega$. These quasilocalized ``5D fields'' describe 4D degrees of freedom localized on the 3-branes, with a penetration depth $\eta$ in the bulk, such that they have no KK-excitations. Note that, in general, each field species (labeled by $i$) has its own WNL length scale $\eta_i$, such that $\forall i \neq j, \ \eta_i \neq \eta_j$. Moreover, since the 5D spacetime symmetries are locally broken to the 4D ones at the fixed point positions, a given smeared field involved in an interaction term on a brane should have a different WNL scale in the directions parallel $\eta_\parallel$ or transverse $\eta_\perp$ to the brane, such that $\eta_\parallel \neq \eta_\perp$. For simplicity, only 1 universal WNL length scale $\eta$ is introduced in this model.

\paragraph{Action:}
The action of the 5D toy model is
\begin{equation}
S_{5D} = \int d^4x \oint dy \left( \mathcal{L}_B + \mathcal{L}_b + \mathcal{L}_{Bb} + \mathcal{L}_{bb} \right) \, , 
\end{equation}
where:
\begin{itemize}[label=$\spadesuit$]
\item $\mathcal{L}_B$ is the bulk Lagrangian of the 5D field $\Phi(x, y)$, with a cubic self-interaction term:
\begin{equation}
\mathcal{L}_B = \dfrac{1}{2} \left[ - \Phi \Delta_\parallel \Phi + \left( \partial_y \Phi \right)^2 \right] + \dfrac{\lambda_B}{3!} \, \widetilde{\Phi}^3 \, ,
\label{Lag_bulk_scalar}
\end{equation}
where $\lambda_B$ is a real coupling (mass dimension $1/2$) that scales as $\sqrt{\Lambda_{UV}}$;
\item $\mathcal{L}_b$ is the free brane Lagrangian of the 4D fields $\theta / \omega(x)$:
\begin{equation}
\mathcal{L}_b = \delta(y) \left( - \dfrac{1}{2} \, \theta \Delta_\parallel \theta \right) + \delta(y - \pi \rho) \left( - \dfrac{1}{2} \, \omega \Delta_\parallel \omega \right) \, ;
\end{equation}
\item $\mathcal{L}_{Bb}$ is the bulk-brane Lagrangian of the interaction terms between the bulk field $\Phi(x, y)$ and the brane fields $\Theta / \Omega(x, y)$:
\begin{equation}
\mathcal{L}_{Bb} = \dfrac{\lambda_{Bb}}{2} \, \widetilde{\Phi} \left( \widetilde{\Theta}^2 + \widetilde{\Omega}^2 \right) \, ,
\label{L_Bb}
\end{equation}
where $\lambda_{Bb}$ is a real coupling (mass dimension $-1/2$) that scales as\footnote{The scaling of the couplings of the brane-localized operators in the WNL model is chosen to match the scaling of the corresponding local model when $\eta \rightarrow 0$, cf. Section~\ref{fuzzy}.} $\eta \sqrt{\Lambda_{UV}}$;
\item $\mathcal{L}_{bb}$ is the brane-brane Lagrangian of the interaction terms between the brane fields $\Theta(x, y)$ and $\Omega(x, y)$:
\begin{equation}
\mathcal{L}_{bb} = \dfrac{\lambda_b}{3!} \left( \widetilde{\Theta}^3 + \widetilde{\Omega}^3 \right) + \dfrac{\lambda_{bb}}{2} \left( \widetilde{\Theta} \, \widetilde{\Omega}^2 + \widetilde{\Omega} \, \widetilde{\Theta}^2 \right) \, ,
\label{L_bb}
\end{equation}
where $\lambda_b$ and $\lambda_{bb}$ are real couplings (mass dimension $-1$) that scales as $\eta^2 \Lambda_{UV}$.
\end{itemize}

\paragraph{Heat Kernel:}
The above Lagrangians appear naively local in terms of the local and smeared fields. To understand why the infinite-derivative feature of the smearing operators introduces WNL, remember that the Gaussian function $\delta_\eta^{(d)}$ is the kernel of the SFT-like smearing operator on $\mathbb{R}^d$, cf. Eq.~\eqref{heat_kernel}. Consider the case where $\eta \ll \rho$, such that one can neglect the compactification effects on the shape of the kernels, and one can use the approximation $\mathbb{E}^5 \simeq \mathbb{R}^5$ as far as only WNL features are concerned. As a consequence, the smeared fields can be expressed as the convolution product:
\begin{equation}
\widetilde{S}(x, y) = \left( \delta_\eta^{(5)} * S \right) (x, y) \, ,
\end{equation}
that is involved in the interaction terms, where WNL is now manifest. In the local limit $\eta \rightarrow 0$, one gets the local fields $\widetilde{S}(x, y) = S (x, y)$ as required, since in the theory of generalized functions \cite{Schwartz:1966}, one has the weak limit:
\begin{equation}
\lim_{\eta \to 0} \delta_\eta^{(5)}(x, y) = \delta^{(5)}(x, y) \, .
\label{single_delta_lim}
\end{equation}

\subsubsection{Kaluza-Klein Dimensional Reduction}
\label{KK_flat_toy}
\paragraph{Local Fields:}
If the couplings are weak, one can perform a perturbative analysis. From the bulk Lagrangian of Eq~\eqref{Lag_bulk_scalar}, the free Euler-Lagrange equation in the bulk is $\Delta \Phi(x, y) = 0$, which is local in spacetime. As usual, one can perform a Kaluza-Klein (KK) decomposition of the 5D field $\Phi(x, y)$:
\begin{equation}
\Phi(x, y) = \sum_{n} \phi_{n}(x) \, f_n(y) \, .
\end{equation}
The 4D fields $\phi_{n}(x)$ describe an infinite tower of KK-modes obeying the 4D Klein-Gordon equations. Each of them has a bulk wave function $f_n(y)$ describing the localization of the KK-mode along the EDS, with Neumann boundary conditions $\partial_y f_n(0, \pi \rho) = 0$. There is a flat 0-mode and an infinite tower of excited modes, with the mass spectrum $m_n = n/\rho$, $n \in \mathbb{N}$, so one can thus define the KK-scale as $M_{KK} = 1 / \rho$. The solutions for the bulk wave functions are
\begin{equation}
f_0(y) = \sqrt{\dfrac{1}{2 \pi \rho}} \, ,
\ \ \ 
\forall n \in \mathbb{N}^*, \ f_n(y) = \sqrt{\dfrac{1}{\pi \rho}} \, \cos \left( \dfrac{n y}{\rho} \right) \, .
\label{local_KK}
\end{equation}

\paragraph{Smeared Fields:}
One can also define a KK-decomposition of the smeared 5D field $\widetilde{\Phi}(x, y)$:
\begin{equation}
\widetilde{\Phi}(x, y) = \sum_{n=0}^{\infty} \widetilde{\phi}_{n}(x) \, \widetilde{f}_n(y) \, ,
\label{KK_smeared}
\end{equation}
with the smeared KK-fields and wave functions:
\begin{equation}
\widetilde{\phi}_n(x) = e^{\eta^2 \partial_\mu^2} \, \phi_n(x) \, , \ \ \ 
\widetilde{f}_n(y) = \left( \delta_\eta * f_n \right)(y) \, ,
\end{equation}
such that
\begin{equation}
\widetilde{f}_0(y) = f_0(y) \, , \ \ \ 
\forall n \in \mathbb{N}^*, \ \widetilde{f}_n(y) = \exp \left[ - \left( \dfrac{n \eta}{\rho} \right)^2 \right] f_n(y) \, .
\label{KK_smeared_wave}
\end{equation}
The mass spectrum of the KK-modes $\widetilde{\phi}_n(x)$ is still $m_n = n/\rho$, and thus ghost-free. Therefore, with $\rho \gg \eta$, only the KK-modes with $n \gg 1$ (i.e. with $m_n \gg \Lambda_\eta$) have their smeared wave functions with a coefficient significantly suppressed with respect to the local case.

\subsubsection{Fuzzy versus $\delta$-like Brane}
\label{fuzzy}

\paragraph{Fuzzy Brane:}
Perhaps the most interesting aspects of WNL in braneworlds, compared to the 4D WNL field theories discussed in the literature, are the new features of fields localized on $\delta$-like branes. Indeed, as the brane-localized fields have their interactions a bit delocalized in the bulk with a penetration depth $\eta$, a $\delta$-like brane does not appear anymore as a singular object in the transverse dimensions but has a small width $\eta$: the terminology of \textit{fuzzy brane} is introduced. Therefore, WNL regularizes the transverse behavior of the branes, and avoids the typical delicate problems of EFT's with $\delta$-like branes. It is instructive to consider the local limit of the interaction terms of the toy model in Section~\ref{toy_model}.

\paragraph{Renormalized Brane Couplings:}
It is useful to remind that
\begin{equation}
\forall N \in \mathbb{N}^*, \ \delta_\eta^N(y) \propto \eta^{1-N} \, \delta_{\eta} \left( \sqrt{N} y \right) \, .
\end{equation}
If a brane-localized operator has a coupling $\alpha(\eta) \propto \eta^{N-1}$, one can define a renormalized coupling $ \alpha_R \in \mathbb{R}$ such that
\begin{equation}
\lim_{\eta \to 0} \alpha(\eta) \, \delta_\eta^N(y) = \alpha_R \, \delta(y) \, .
\end{equation}
In the following, one can thus formally perform the following replacement:
\begin{equation}
\forall N \in \mathbb{N}^*, \ \delta_\eta^N(y)  \underset{\eta \rightarrow 0}{\mapsto} \delta(y) \, ,
\label{delta_power}
\end{equation}
with a renormalization of the coupling of the corresponding brane-localized operator. One can check that the couplings of the toy models considered in this article scale with powers of $\eta$ which are consistent with this discussion on renormalized couplings in the local limit $\eta \rightarrow 0$. In the special case of operators involving fields localized on different fuzzy branes (which have a trivial local limit), the scaling can be checked by considering an analogous operator involving these fields localized on the same fuzzy brane, and then by taking the local limit.

\paragraph{Brane-Bulk Interactions:}
From Eq.~\eqref{L_Bb}, one can write the interaction terms between bulk and brane fields as
\begin{equation}
\mathcal{L}_{Bb} = \dfrac{\lambda_{Bb}}{2} \, \widetilde{\Phi} \left[ \delta_\eta^2(y) \, \widetilde{\theta}^2 + \delta_\eta^2(y-\pi\rho) \, \widetilde{\omega}^2 \right] \, .
\label{L_Bb_2}
\end{equation}
By integrating over the EDS, one gets the effective 4D Lagrangian of the KK-modes:
\begin{equation}
\oint dy \ \mathcal{L}_{Bb} = \sum_{n=0}^\infty \dfrac{\lambda_{Bb}^{(n)}}{2} \, \widetilde{\phi}_n \left( \widetilde{\theta}^2 + \widetilde{\omega}^2 \right) \, ,
\end{equation}
with the effective 4D couplings (mass dimension 1):
\begin{equation}
\forall n \in \mathbb{N}, \ \lambda_{Bb}^{(n)} = \lambda_{Bb} \oint dy \ \widetilde{f}_n(y) \, \delta_\eta^2(y) \, .
\end{equation}
From Eqs.~\eqref{heat_kernel} and \eqref{KK_smeared_wave}, one gets (with a natural coupling $\Lambda_{Bb} \sim \eta \sqrt{\Lambda_{UV}}$):
\begin{align}
\lambda_{Bb}^{(0)} \sim \dfrac{1}{4 \pi} \sqrt{\dfrac{\Lambda_{UV}}{\rho}} \, , \ \ \ 
\forall n \in \mathbb{N}^*, \ \lambda_{Bb}^{(n)} \sim \dfrac{1}{2 \pi} \sqrt{\dfrac{\Lambda_{UV}}{2 \rho}} \, \exp \left[ - \dfrac{3}{2} \left( \dfrac{n \eta}{\rho} \right)^2 \right] \, .
\end{align}
Only the couplings $\lambda_{Bb}^{(n)}$ to KK-modes with $m_n \gg \Lambda_\eta$ are significantly suppressed via WNL. If one takes the local limit $\eta \rightarrow 0$ in Eq.~\eqref{L_Bb_2} and uses the formal replacement in Eq.~\eqref{delta_power}, one gets
\begin{equation}
\mathcal{L}_{Bb} \underset{\eta \rightarrow 0}{\propto} \Phi \left[ \delta(y) \, \theta^2 + \delta(y-\pi\rho) \, \omega^2 \right] \, ,
\end{equation}
which is exactly the form of an interaction term in a local braneworld EFT with a $\delta$-like brane.

\paragraph{Brane-Brane Interactions:}
From Eq.~\eqref{L_bb}, one can write the interaction terms between brane fields as
\begin{equation}
\mathcal{L}_{bb} = \dfrac{\lambda_{b}}{3!} \left[ \delta_\eta^3(y) \, \widetilde{\theta}^3 + \delta_\eta^3(y-\pi\rho) \, \widetilde{\omega}^3 \right]
+ \dfrac{\lambda_{bb}}{2} \left[ \delta_\eta(y) \, \delta_\eta^2(y-\pi\rho) \, \widetilde{\theta} \widetilde{\omega}^2 + \delta_\eta(y-\pi\rho) \, \delta_\eta^2(y) \, \widetilde{\omega} \widetilde{\theta}^2 \right] \, .
\label{L_bb_2}
\end{equation}

\subparagraph{Same Brane:}
One can take $\lambda_{b} \neq 0$ and $\lambda_{bb} = 0$ to discuss the first term, which can be rewritten as
\begin{equation}
\oint dy \ \mathcal{L}_{bb} = \oint dy \ \mathcal{L}_{bb}^\prime \, ,
\end{equation}
where
\begin{equation}
\mathcal{L}_{bb}^\prime = \dfrac{\lambda_{b}^\prime}{3!} \left[ \delta(y) \, \widetilde{\theta}^3 + \delta(y-\pi\rho) \, \widetilde{\omega}^3 \right] \, ,
\label{bb_local}
\end{equation}
with the effective 4D coupling (mass dimension 1):
\begin{equation}
\lambda_{b}^\prime = \lambda_{b} \oint dy \ \delta_\eta^3(y) \sim \sqrt{\dfrac{1}{3}} \, \dfrac{\Lambda_{UV}}{4 \pi} \, ,
\label{rescale_same_brane}
\end{equation}
by using Eq.~\eqref{heat_kernel}, and with a natural coupling $\lambda_b \sim \eta^2 \Lambda_{UV}$. Therefore, for an interaction term between 4D degrees of freedom localized on the same fuzzy brane, the effect of WNL along the EDS is just to rescale\footnote{However, WNL cannot be used to generate a suppressed coupling with respect to its natural value for interacting fields localized on the same fuzzy brane, since the rescaling factor in Eq.~\eqref{rescale_same_brane} is not very small.} the coupling constant of the corresponding brane operator with respect to the same interaction in a local model with a $\delta$-like brane. Indeed, if one takes the local limit $\eta \rightarrow 0$ in Eq.~\eqref{L_bb_2} with the formal replacement in Eq.~\eqref{delta_power}, one obtains
\begin{equation}
\mathcal{L}_{bb} \underset{\eta \rightarrow 0}{\propto} \left[ \delta(y) \, \theta^3 + \delta(y-\pi\rho) \, \omega^3 \right] \, ,
\end{equation}
whose localized fields have the same singular transverse features as in the Lagrangian of Eq.~\eqref{bb_local}.\\

\subparagraph{Different Branes:}
One considers $\lambda_{b} = 0$ and $\lambda_{bb} \neq 0$ to discuss the second term in Eq.~\eqref{L_bb_2}, such that
\begin{equation}
\oint dy \ \mathcal{L}_{bb} = \dfrac{\lambda_{bb}^\prime}{2} \left( \widetilde{\theta} \widetilde{\omega}^2 + \widetilde{\omega} \widetilde{\theta}^2 \right) \, ,
\end{equation}
with the effective 4D coupling (mass dimension 1):
\begin{align}
\lambda_{bb}^\prime = \lambda_{bb} \oint dy \ \delta_\eta(y) \, \delta_\eta^2(y-\pi\rho)
\sim \sqrt{\dfrac{1}{3}} \, \dfrac{\Lambda_{UV}}{4 \pi} \, \exp \left[ - \dfrac{1}{6} \left( \dfrac{\pi \rho}{\eta} \right)^2 \right]
\underset{\rho \gg \eta}{\ll} 1 \, ,
\end{align}
by using Eq.~\eqref{heat_kernel}, and with a natural coupling $\lambda_{bb} \sim \eta^2 \Lambda_{UV}$. Therefore, when $\rho \gg \eta$, it is possible to get naturally suppressed couplings between the fields localized on the different fuzzy branes: the transverse smearing kernels $\delta_\eta(y)$ and $\delta_\eta(y-\pi \rho)$ have a tiny overlap. Note that in the local limit $\eta \rightarrow 0$, one has $\lambda_{bb}^\prime \rightarrow 0$, and one recovers that fields localized on different $\delta$-like branes do not couple directly in a local EFT: one needs a bulk mediator field for that purpose. Another way to obtain this result is by taking the local limit $\eta \rightarrow 0$ directly in Eq.~\eqref{L_bb_2} with the formal replacement in Eq.~\eqref{delta_power}, such that
\begin{equation}
\mathcal{L}_{bb} \underset{\eta \rightarrow 0}{\propto} \delta(y) \, \delta(y-\pi\rho) \left( \theta \omega^2 + \omega \theta^2 \right) = 0 \ \ \ \text{since} \ \ \ 
\delta(y) \, \delta(y-\pi\rho) = 0 \, ,
\end{equation}
because these 2 Dirac generalized functions have disjointed pointlike supports ($\rho > 0$).

\subsection{Weakly Nonlocal Gauge Fields}
\subsubsection{4D Case in a Nutshell}
\label{4Dgauge_intro}
The difficulty for UV finiteness appears for gauge symmetries, since it is not possible to smear interaction terms independently of the kinetic terms \cite{Chretien:1954we}. The manifest gauge invariant scheme proposed by Krasnikov \cite{Krasnikov:1987yj} (where gauge-covariant derivatives appear in the WNL form factors) is the one that is used in string-inspired nonlocal models \cite{Biswas:2014yia, Ghoshal:2017egr, Ghoshal:2020lfd, Frasca:2020jbe, Frasca:2020ojd, Frasca:2021iip, Mo:2022szw, Chatterjee:2023ehr}. For instance, a string-inspired nonlocal Yang-Mills action has the form
\begin{equation}
S_{YM} = \dfrac{1}{2 g^2} \int d^dx \ \text{Tr} \left[ F_{\mu \nu} \, e^{- \eta^2 D_\mu^2} F_{\mu \nu} \right] \, ,
\end{equation}
where $D_\mu$ is the 4D gauge-covariant derivative, $F_{\mu \nu}$ is the 4D gauge-covariant field strength, and $g$ is the 4D gauge coupling constant. Such a theory is not UV finite because there is a competition between the WNL form factors in the kinetic and interaction terms. For SFT-like form factors, the perturbative renormalization program fails\footnote{Even if 1-loop computations naively led to a UV scale invariant behavior \cite{Ghoshal:2017egr, Ghoshal:2020lfd}, these QFT's violate Weinberg's power counting theorem \cite{Weinberg:1959nj}: higher-loop 1PI diagrams are not under perturbative control and these WNL-QFT's are not renormalizable by usual perturbative techniques \cite{Talaganis:2014ida, Buchbinder:2021wzv}.}: the problem is that one wants to treat the nonlocal form factor nonperturbatively while integrating-out the full tower of string excitations of the stringy UV completion that introduces UV/IR mixing \cite{Abel:2021tyt, Abel:2023hkk}. One can thus use the nonlocal action to determine background solutions in a semiclassical analysis. However, it is necessary to expand perturbatively the form factor in order to have a meaningful local QFT below the string scale ($\Lambda_\eta \sim M_s$):
\begin{equation}
e^{- \eta^2 D_\mu^2} \simeq \sum_{k=0}^{K-1} \dfrac{1}{k!} \left( - \dfrac{D_\mu^2}{\Lambda_{\eta}^2} \right)^k + \mathcal{O} \left( \dfrac{1}{\Lambda_{\eta}^{2K}} \right) \, ,
\end{equation}
that can be handled as a usual EFT. The application to gravity follows an analogous covariant scheme \cite{Biswas:2014tua}.

\subsubsection{Compactified Extra Dimension}
In the original braneworld models, the SM are 4D degrees of freedom localized on a 3-brane \cite{Sundrum:1998sj}. In string model building \cite{Becker:2007zj, Ibanez:2012zz}, D-brane stacks (where open strings are attached) realize gauge symmetries in their worldvolume. In the following examples about WNL gauge theories, some stringy UV completion is assumed: only SFT-like smearing effects are taken into account, and the extra modifications due to the Dirac-Born-Infeld nature of a brane effective action are ignored.

\paragraph{Brane-Localized Gauge Symmetries:}
Consider a toy 5D model with a flat EDS and a 3-brane at $y=0$, where are localized: (i) $SU(N_C)$ gauge bosons\footnote{The case of an Abelian $U(1)$ gauge field can be deduced easily form the non-Abelian example.} $A_\mu^a(x)$ with $a \in \llbracket 1, N_C \rrbracket$ and the WNL length scale $\eta_g$; (ii) $N_F$ flavor of scalar fields $\phi_i(x)$ in the fundamental representation of $SU(N_C)$ with $i \in \llbracket 1, N_F \rrbracket$ and the associated WNL length scales $\eta_i$. The brane action describing the minimal gauge interactions is (up to commutators of gauge-covariant derivatives):
\begin{equation}
\int dy \, dx \left\{ \dfrac{1}{2 g^2} \, \text{Tr} \left[ F_{\mu \nu} \, e^{\eta_g^2 \left( \partial_y^2 - D_\mu^2 \right)} F_{\mu \nu} \right] + \sum_{i=1}^{N_F} \left( D_\mu \phi_i \right)^\dagger e^{\eta_i^2 \left( \partial_y^2 - D_\mu^2 \right)} D_\mu \phi_i \right\} \delta(y) \, .
\label{action_gauge_1}
\end{equation}
The breaking of Lorentz-Poincaré symmetries by the brane allows choosing independent longitudinal and transverse form factors for the same field. Since we deal with a brane-localized 4D gauge invariance, the transverse smearing of the brane does not involve gauge-covariant derivatives, and one has normalized Gaussian profiles:
\begin{equation}
\int dy \ e^{\eta_g^2 \partial_y^2} \delta(y) = \int dy \ e^{\eta_i^2 \partial_y^2} \delta(y) = 1 \, .
\end{equation}
Therefore, even if the $N_F$ flavors of scalar fields are smeared by a different WNL length scale $\eta_i$, the coupling to brane-localized gauge bosons is flavor-blind (as in the SM) without invoking some extra symmetry principle.

\paragraph{Bulk Gauge Symmetries:}
Higher-dimensional generalization of string-inspired gauge invariant models is straightforward from the 4D literature. For instance, consider again a 5D model with a flat EDS, where the $SU(N_F)$ gauge bosons $\mathcal{A}_\mu^a(x,y)$ minimally coupled to the $N_F$ scalar fields $\Phi_i(x,y)$ propagate into the bulk. The action is
\begin{equation}
\int dy \, dx \left\{ \dfrac{1}{2 g^2} \, \text{Tr} \left[ \mathcal{F}_{MN} \, e^{-\eta_g^2 \mathcal{D}_M^2} \mathcal{F}_{MN} \right] + \sum_{i=1}^{N_F} \left( D_M \Phi_i \right)^\dagger e^{-\eta_i^2 \mathcal{D}_M^2} D_M \Phi_i \right\} \, ,
\label{action_gauge_2}
\end{equation}
where $\mathcal{D}_M$ is the 5D gauge-covariant derivative, $\mathcal{F}_{MN}$ is the 5D gauge-covariant field strength, and $g$ is the 5D gauge coupling constant. Here, 5D Lorentz-Poincaré and gauge invariance impose the same WNL form factor for all components of a 5D field and a flavor blind gauge coupling.

\paragraph{Brane-Bulk Gauge Interactions:}
The last possibility is if the 5D scalar fields $\Phi_i(x,y)$ of the action in Eq.~\eqref{action_gauge_2} are replaced by 4D scalars $\phi_i(x)$ localized on a 3-brane at $y=0$, such as
\begin{equation}
\int dy \, dx \left\{ \dfrac{1}{2 g^2} \, \text{Tr} \left[ \mathcal{F}_{MN} \, e^{-\eta_g^2 \mathcal{D}_M^2} \mathcal{F}_{MN} \right] +
\sum_{i=1}^{N_F} \left[ \left( D_\mu \phi_i \right)^\dagger e^{\eta_i^2 \left( D_y^2 - D_\mu^2 \right)} D_\mu \phi_i \right] \delta(y) \right\} \, ,
\end{equation}
where brane-localized kinetic terms for the bulk gauge fields \cite{Carena:2002me} are not considered for simplicity.
The difference with the brane action in Eq.~\ref{action_gauge_1} is the 5D gauge symmetry that requires the dressing of the brane-localized charge fields by a cloud of gauge bosons that becomes relevant at the scales $\eta_i$.

\subsubsection{Non-Stringy UV Completions}
\label{beyond_strings}
This last decade, another class of WNL field theories has received more attention: these QFT's realize superrenormalizability or UV finiteness in pure gravity and gauge theories in any number of spacetime dimensions \cite{Kuzmin:1989sp, Tomboulis:1997gg, Modesto:2011kw, Modesto:2013oma, Modesto:2014lga, Tomboulis:2015esa, Modesto:2015lna, Modesto:2015foa, Modesto:2017sdr, BasiBeneito:2022wux}. They interpolate between 2 local QFT's: a 2-derivative one in the IR and a higher-derivative one in the UV. In order to avoid ghost-like degrees of freedoms, the interpolating region is a nonlocal window: the infinite-derivative form factors must be asymptotically polynomial in the UV for both imaginary and real energies. Instead of the SFT-like form factors, they allow the quantum corrections to a WNL gauge theory to be under perturbative control. Nevertheless, a ghost-free Higgs mechanism is a more subtle issue \cite{Gama:2018cda, Hashi:2018kag}. For that purpose, a recent recipe \cite{Modesto:2021ief} allows building theories which have the same spectrum and tree level scattering amplitudes than their local limit \cite{Modesto:2021soh}, such that WNL manifests only at loop level and one can build a ghost-free Higgs mechanism \cite{Modesto:2021okr}. In this paper, we will not discuss further these theories, but the interested reader could easily generalize the discussion to other WNL theories.

\subsection{Shadow Extra Dimensions}
What happens if one wants to trust the WNL field theory above the WNL scale? In Section~\ref{4Dgauge_intro}, it was already discussed that for SFT-like form factors, gauge theories are not under perturbative control in this simplified framework. The following more speculative section will focus on a scalar field that does not have this problem, and the case of the gauge theories with a nonlocal window in Section~\ref{beyond_strings} will be briefly mentioned.\\

In order to have a taste of the phenomenology of the KK-excitations with WNL, it is useful to consider the smeared 4D fields $\widetilde{\phi}_n(x)$ for the KK-modes, which appear in the WNL interaction terms of the EFT after dimensional reduction to 4D. In this class of WNL field theories, it is possible to rewrite the Lagrangians and Euler-Lagrange equations only in terms of the smeared fields \cite{Buoninfante:2018mre}, such that WNL then appears only in the kinetic terms of the smeared KK-fields:
\begin{equation}
- \dfrac{1}{2} \, \widetilde{\phi}_n \, e^{-2 \eta^2 \partial_\mu^2} \left( \partial_\mu^2 - m_n^2 \right) \widetilde{\phi}_n \, ,
\end{equation}
such that one can easily extract their propagators:
\begin{equation}
\Pi_n(p^2) = \dfrac{-i e^{-2 \eta^2  p^2}}{p^2 + m_n^2} \, ,
\end{equation}
which are exponentially suppressed in the UV, i.e. when $p^2 > \Lambda_\eta^2$, reflecting the UV opacity of such WNL field theories. Moreover, in Section~\ref{fuzzy}, it was shown that these KK-modes have also suppressed couplings with brane fields. Therefore, any contribution to a process from a KK-particle whose mass is $m_n > \Lambda_\eta$ will be suppressed compared to a local model\footnote{Note that the status of such transnonlocal states is not clear, and it was suggested in Ref.~\cite{Buoninfante:2018mre} that no external states can be associated to them.}.\\

Nevertheless, once the $S$-matrix elements are computed, there is the issue of the Efimov analytic continuation of the external momenta to real energies \cite{Efimov:1965mnl, Efimov:1966ylf, Pius:2016jsl, DeLacroix:2018arq, Carone:2016eyp, Briscese:2018oyx, Chin:2018puw, Briscese:2021mob, Koshelev:2021orf, Buoninfante:2022krn}. Then, the SFT-like smearing operator of the toy model considered here is known to have a strong coupling problem for energies above $\Lambda_\eta$: $S$-matrix elements blow up at high center of mass energy in the WNL regime \cite{Chin:2018puw, Koshelev:2021orf}. Nevertheless, it was mentioned in Ref.~\cite{Pius:2016jsl} that in a realistic SFT-derived model (well beyond this toy model), there is compensation between different vertices such that this problem does not appear. Moreover, the qualitative results of this article should not be affected if one takes another WNL form factor of the literature (beyond SFT), which gives UV suppression in both real and imaginary time directions \cite{Chin:2018puw, Koshelev:2021orf}, or a gauge theory with a nonlocal window that exhibits a softened UV behavior \cite{Kuzmin:1989sp, Tomboulis:1997gg, Modesto:2011kw, Modesto:2013oma, Modesto:2014lga, Tomboulis:2015esa, Modesto:2015lna, Modesto:2015foa, Modesto:2017sdr, BasiBeneito:2022wux}.\\

As a consequence, in a realistic model, the effects of such KK-excitations will be much more difficult to probe in an experiment. In full analogy with the discussion in Ref.~\cite{Biswas:2014yia}, the experimenters will see that the cross-sections predicted in the SM are suppressed for energies $E > \Lambda_\eta$, concluding to a WNL UV completion of the SM. However, they will have more difficulties observing the KK-tower if $\eta \sim \rho$, and then concluding the existence of the EDS. Note that in this particular case, WNL smears the brane all along the EDS, such that the difference between bulk and brane fields is meaningless: all poles of the KK-excitations are in the UV opaque regime, and one should better understand this UV regime of WNL-QFT's to be able to really discuss the phenomenology of such \emph{shadow} EDS. Moreover, in the case of a UV completion in string theory, where $M_s \sim \Lambda_\eta$, one expects that Regge excitations also appear in the WNL regime, and one should use the full stringy UV completion to study the phenomenology.

\section{Bottom-Up Model Building \& Applications}
\label{applications}

\subsection{Flavor Hierarchy from Fuzzy Branes}
\label{smeared_hierarchy}

\subsubsection{Split Fermions in a Nutshell}
In the SM of particle physics, the observed fermion mass spectrum \cite{Workman:2022ynf} needs to introduce a very hierarchical pattern for the Yukawa couplings. From a Dirac naturalness point of view, one can wonder why the Yukawa couplings are so hierarchical, and why the CKM matrix appears so close to the identity, since no particular texture is preferred in the SM. A related issue is the smallness of the neutrino masses (of Dirac/Majorana type), which involves another hierarchy with the charged leptons.\\

Among all the creative proposals to solve this flavor puzzle in the literature, one of them is the split fermion scenario, originally proposed by Arkani-Hamed and Schmaltz (AS) in Ref.~\cite{ArkaniHamed:1999dc}, with the first realistic model to solve the flavor puzzle in Ref.~\cite{Mirabelli:1999ks}. The central idea behind the AS proposal (cf. Ref.~\cite{ArkaniHamed:1999dc} for details) is that the different species of SM fermions are ``stuck'' at different points along at least 1 (flat) EDS, with the SM gauge and Higgs fields identified with the flat zero mode of bulk fields. In this way, the 4D effective Yukawa couplings are suppressed by the tiny overlap between the wave functions along the EDS of the 2 chiral fermions and the Higgs field. In the original AS model, the fermion zero modes are localized by a background scalar field with a kink profile along the EDS, i.e. a domain wall.\\

The aim of this Section~\ref{split_fuzzy} is to realize the AS idea, but without the need for a domain wall in the 5D EFT to trap the chiral fermions. Each 4D fermion is now localized on a different fuzzy 3-brane, such that they are delocalized in the bulk by a Gaussian smearing form factor originating from WNL. In the proposed class of WNL models, one considers a 5D Euclidean spacetime $\mathbb{E}^5 = \mathbb{R}^4 \times [0, \pi \rho]$, where the gauge and Higgs fields propagate in the bulk.

\subsubsection{Multiple Fuzzy Branes}
\label{split_fuzzy}

\paragraph{Toy Model:}
In order to illustrate the above idea, it is enough to consider a toy model with a real 5D Higgs-like scalar field $H(x, y)$ with a potential $V(H)$, and Yukawa couplings to 2 brane-localized Weyl fermions of opposite chirality, i.e. a 4D left/right-handed fermion $\psi_{L/R}(x)$ localized on a 3-brane at $y = y_{L/R}$. All these fields have a mass dimension $3/2$. As we will see, the crucial feature of the model is that the Weyl fermions are smeared in the bulk. Therefore, it is enough to assume that the WNL scales associated to the fermion fields are lower than the one associated to $H(x, y)$ in the bulk. Then, one can consider a 5D EFT where the WNL features of the Higgs-like field decouple, and only the 4D fermion fields are smeared by WNL, with a universal WNL length scale $\eta$ for simplicity.

\paragraph{Action:}
The smeared fields $\widetilde{\Psi}_{L/R}(x, y)$ are defined in terms of the 5D localized fields $\Psi_{L/R}(x, y) = \psi_{L/R}(x) \, \delta(y - y_{L/R})$, such that
\begin{align}
\widetilde{\Psi}_{L/R}(x, y)
= e^{\eta^2 \Delta} \, \Psi_{L/R}(x, y)
= \widetilde{\psi}_{L/R}(x) \, \delta_\eta(y - y_{L/R}) \, ,
\end{align}
with
\begin{align}
\widetilde{\psi}_{L/R}(x) = e^{\eta^2 \Delta_\parallel} \, \psi_{L/R}(x) \, , \ \ \ 
\delta_\eta(y - y_{L/R}) = e^{\eta^2 \Delta_\perp} \, \delta(y - y_{L/R}) \, ,
\end{align}
where $\delta_\eta(y)$ is well approximated by a 1D Gaussian function in Eq.~\eqref{heat_kernel} if $\eta \ll \rho$, and the brane positions are sufficiently far from the EDS boundaries. The width of these Gaussian profiles along the EDS is given by the WNL length scale $\eta$, which plays the same role as the domain wall width in the original AS model \cite{ArkaniHamed:1999dc}. The 5D action is
\begin{equation}
S_{5D} = \int d^4x \int_0^{\pi \rho} dy \ \left( \mathcal{L}_{H} + \mathcal{L}_{L} + \mathcal{L}_{R} + \mathcal{L}_{Y} \right), 
\end{equation}
with the Lagrangians
\begin{align}
&\mathcal{L}_H = \dfrac{1}{2} \left[ - H \Delta_\parallel H + \left( \partial_y H \right)^2 \right] - V(H) \, , \nonumber \\
&\mathcal{L}_{L/R} = \delta(y - y_{L/R}) \, \psi_{L/R}^\dagger (-i \cancel{\partial}) \psi_{L/R} \, , \nonumber \\
&\mathcal{L}_{Y} = - Y \, i \widetilde{\Psi}_L^\dagger H \widetilde{\Psi}_R + \text{H.c.} \, ,
\label{Yukawa_Lag_1}
\end{align}
where the real 5D Yukawa coupling $Y$ has a mass dimension $-3/2$ and scales as $\eta \sqrt{\ell_{UV}}$, where $\ell_{UV}$ is some UV length scale defined as the EFT UV cutoff in the local limit $\eta \rightarrow 0$.

\paragraph{Effective 4D Hierarchy:}
By performing a KK dimensional reduction, the 5D Higgs-like field can be decomposed around its flat vacuum expectation value (VEV) $v$ in 4D KK-modes $h_n(x)$ (mass dimension 1), such that
\begin{equation}
H(x, y) = \dfrac{v}{\sqrt{\pi \rho}} + \sum_n h_n (x) \, f_n(y) \, ,
\end{equation}
where $v$ is a mass scale, the $f_n(y)$'s are the KK wave functions (mass dimension $1/2$) associated to the KK-modes $h_n(x)$. The EDS is flat, so the KK-scale is defined as $M_{KK} = 1/\rho$. As in the original AS model, $V(H)$ is chosen such that the 0-mode (identified with a SM-like Higgs boson) has a flat wave function:
\begin{equation}
f_0(y) = \sqrt{\dfrac{1}{\pi \rho}} \, .
\label{zero_h}
\end{equation}
In the 4D EFT obtained by integrating over the EDS, the action includes the terms:
\begin{equation}
S_{4D} \supset -i \int d^4x \ \psi^\dagger \left( \cancel{\partial} + m_\psi + y_0 \, h_0 \right) \psi \, ,
\end{equation}
where $m_\psi = y_0 \, v$ is the mass of the Dirac fermion
\begin{equation}
\psi (x) =
\begin{pmatrix}
\psi_L (x) \\
\psi_R (x)
\end{pmatrix} \, ,
\end{equation}
and the effective 4D Yukawa coupling $y_0$ (mass dimension 0) between $\psi(x)$ and $h_0(x)$ is given by the overlap:
\begin{equation}
y_0 = Y \int_0^{\pi \rho} dy \ f_0(y) \, \delta_\eta (y-y_L) \, \delta_\eta (y-y_R) \, .
\end{equation}
By assuming a natural 5D Yukawa coupling $Y \sim \eta \sqrt{\ell_{UV}}$, and an interbrane distance $r = |z_L - z_R| \gg \eta$, one gets
\begin{equation}
y_0 \sim \dfrac{1}{2 \pi} \sqrt{\dfrac{\ell_{UV}}{2 \rho}} \, \exp \left[ - \dfrac{1}{8} \left( \dfrac{r}{\eta} \right)^2 \right] \ll 1 \, ,
\end{equation}
by using Eqs.~\eqref{heat_kernel} and \eqref{zero_h}. Therefore, $y_0$ is naturally exponentially suppressed for sufficiently separated fuzzy branes.

\paragraph{Flavor Hierarchy:}
Since the transverse Gaussian kernels of the fermions have the same features as the bulk wave functions of the trapped fermions in the original AS model \cite{ArkaniHamed:1999dc}, one can localize the fermions at the same EDS points and reproduce the same mass matrix and mixing angles as in Ref.~\cite{Mirabelli:1999ks} to explain the SM Yukawa hierarchy. For energies below the WNL scale $\Lambda_\eta$, the phenomenology of this WNL model is completely identical to an AS one, so one has the same constraints as in Ref.~\cite{Mirabelli:1999ks}. One needs to probe the scale associated to the transverse Gaussian fermion profiles to resolve the microscopic details of the model, e.g. fermions trapped inside a domain wall or an intrinsic nonlocal UV theory (e.g. open strings attached to D-branes). Note that for other WNL form factors, the transverse kernels are not Gaussian anymore but are still sharply localized, so this qualitative discussion still holds.

\paragraph{Intersecting Branes:}
In top-down constructions, another interesting scenarios are semirealistic string models with branes wrapping compact cycles that intersect at nontrivial angles \cite{Berkooz:1996km}: such models were intensively studied in string phenomenology \cite{Ibanez:2012zz}. They give rise to several copies of chiral fermions localized at the intersections, thus they offer a natural explanation of the SM fermion generations. The computations of Yukawa couplings are done in string theory. Therefore, the WNL framework offers for the first time an EFT description of flavor models with branes intersecting at angles. Indeed, in a local EFT with $\delta$-like branes, one cannot describe Yukawa interactions between fields localized on different branes without matching contact interactions with a full stringy computation (e.g. Refs.~\cite{Abel:2003fk, Chemtob:2008cb}). The approach proposed here allows a more bottom-up approach to model building with intersecting brane, since the use of an auxiliary domain wall to localize the 0-modes is not fair from the point of view of the UV completion.

\subsection{Scale Hierarchy from a String-Inspired Warped Throat}
\label{warped_transmut}

\subsubsection{Motivations}
In an EFT with a UV cutoff $\Lambda_{UV}$, the existence of an elementary Higgs-like scalar field of mass $m \ll \Lambda_{UV}$ is expected to require fine-tuning in the UV completion, if the limit $m \rightarrow 0$ does not enhance the number of symmetries of the action, which is known as the gauge hierarchy problem \cite{Hebecker:2020aqr}. WNL field theories give a potential solution to this issue \cite{Krasnikov:1987yj, Moffat:1990jj, Biswas:2014yia}: the radiative corrections to the mass of the $H^0$ boson scale as $\delta M_H^2 \propto \Lambda_\eta^2$, instead of $\delta M_H^2 \propto \Lambda_{UV}^2$ in the local SM. A light electroweak (EW) Higgs sector is then natural if $\Lambda_\eta$ is at the terascale, even if a larger new scale like the Planck scale $\Lambda_P \gg \Lambda_\eta \gtrsim M_H$ exists\footnote{Other higher-derivative field theories with similar results are Lee-Wick (LW) theories \cite{Grinstein:2007mp, Espinosa:2007ny, Espinosa:2011js}: $\delta M_H^2 \sim M_{LW}^2$, where $M_{LW}$ is the mass of the first LW-partner.}. Nevertheless, it does not explain the gauge hierarchy $\Lambda_P \gg \Lambda_\eta$ without extra ingredients.\\

The paradigm of warped EDS's, originally proposed by Randall and Sundrum (RS) in Ref.~\cite{Randall:1999ee}, offers a powerful scenario to considerably reduce the fine-tuning issue of a light scalar, like the $H^0$ boson. In the original RS model (RS1) \cite{Randall:1999ee}, the Higgs field of the EW theory is a 4D field localized at the IR boundary (IR-brane) of a slice of a 5D anti-de Sitter (AdS$_5$) spacetime (RS throat). The scale $\Lambda_{IR}$, at which gravity becomes strongly coupled on the IR-brane, and the VEV of the EW Higgs field are exponentially redshifted from the 5D gravity scale $\Lambda_{UV} \sim \Lambda_P$ on the UV-brane, by the AdS$_5$ warp factor: one could talk about warp transmutation of scales. One needs $\Lambda_{IR}$ at the terascale to avoid unnatural fine-tuning.\\

Attempts to UV complete the RS-like models in string theory exist, e.g. with warped throat geometries à la Klebanov-Strassler \cite{Klebanov:2000hb}, but it remains very challenging to find a fully realistic construction \cite{Reece:2010xj}. Note that in such string theory models, the geometry near the IR-tip of the throat in not AdS$_5$, but this does not spoil the RS mechanism of warp transmutation of scales. After integrating out the Regge excitations, stringy effects can be encapsulated into the 5D EFT through higher-dimensional operators, and the extended features of the strings remain through WNL form factors. The aim of the Section~\ref{transmut_brane} is to show that a warped EDS realizes the warp transmutation of both the brane scalar mass and the WNL scale\footnote{For another mechanism of WNL scale transmutation without EDS's, cf. Ref.~\cite{Buoninfante:2018gce}.}. In order to illustrate the mechanism, it is enough to focus on a toy model with a real Higgs-like scalar field in a SFT-inspired WNL field theories, and it is straightforward to generalize the discussion to more realistic WNL Higgs-like models.

\subsubsection{Warp Transmutation of Scales}
\label{transmut_brane}

\paragraph{Geometrical Background:}
Let a model with a warped EDS compactified on the orbifold $S^1/\mathbb{Z}_2$ of proper length $\pi \rho$, i.e. a 5D Euclidean spacetime with a metric as in Eq~\eqref{warped_metric}. The boundaries at $y = 0$ and $y = \pi \rho$ are called UV- and IR-branes, respectively. It is assumed that this kind of background can arise from some WNL extension of Einstein-Hilbert gravity. The 5D reduced Planck scale, which is also the UV cutoff on the UV-brane, is noted $\Lambda_{UV}$.

\paragraph{IR-Brane Field:}
A 4D real scalar field $H(x)$ (mass dimension 1) is localized on the IR-brane with a quartic self-interaction. Since it is a brane-brane interaction, it is enough to consider a smeared field $\widetilde{H}(x)$, which is delocalized only in the directions parallel to the IR-brane:
\begin{equation}
\widetilde{H}(x) = e^{\eta^2 \Delta_\parallel} \, H(x) \, .
\end{equation}
Indeed, as discussed in Section~\ref{fuzzy}, a delocalization transverse to the brane is equivalent to a rescaling of the self-coupling $\lambda \in \mathbb{R}_+$ (mass dimension 0). The brane action is
\begin{equation}
S_H = \int d^4x \oint dy \ \delta (y-\pi \rho) \, \sqrt{g} \, \left[ - \dfrac{1}{2} \, H \left( \Delta_\parallel - \mu_H^2 \right) H + \dfrac{\lambda}{4!} \, \widetilde{H}^4 \right] \, ,
\label{H_Lag}
\end{equation}
with the mass parameter $\mu_H^2 \in \mathbb{R}$. $S_H$ is invariant under the $\mathbb{Z}_2$ transformation $H(x) \mapsto -H(x)$. When $\mu_H^2 \geq 0$, $H(x)$ is a Klein-Gordon field (vanishing VEV), whereas for $\mu_H^2 < 0$, it has a nonvanishing VEV that spontaneously breaks the $\mathbb{Z}_2$ symmetry, like a Higgs field for a continuous gauge symmetry. The choice of a SFT-like form factor is not important for the qualitative statements of this section, e.g. one can instead consider another UV damped form factor.

\paragraph{Bulk Field:}
A 5D scalar field $\Phi(x, y)$ (mass dimension $3/2$) propagates all along the EDS, and interacts with the brane field $H(x)$, such that the 5D action is
\begin{equation}
S_{5D} = \int d^4x \oint dy \ \sqrt{g} \left[ \mathcal{L}_{B} + \delta (y - \pi \rho ) \, \mathcal{L}_{b} \right] \, ,
\end{equation}
with the bulk Lagrangian
\begin{equation}
\mathcal{L}_B = \dfrac{1}{2} \left[ - \Phi \Delta_\parallel \Phi + \left( \partial_y \Phi \right)^2 \right] \, ,
\end{equation}
and the brane Lagrangian
\begin{equation}
\mathcal{L}_{b} = \lambda_{\Phi H} \, \widetilde{\mathcal{O}}_\Phi \, \widetilde{\mathcal{O}}_H \, .
\end{equation}
The brane-localized generic operators $\widetilde{\mathcal{O}}_\Phi$ and $\widetilde{\mathcal{O}}_H$ involve the smeared fields
\begin{equation}
\widetilde{\Phi}(x, y) = e^{\eta^2 \Delta_\parallel} \, \Phi (x, y)
\end{equation}
and $\widetilde{H}(x)$, respectively, and the coupling $\lambda_{\Phi H} \in \mathbb{R}$ is weak. Note that for simplicity, the fields are smeared only in the directions parallel to the IR-brane, and this choice is allowed by the spacetime symmetries at the brane position. Considering instead a sufficiently narrow fuzzy IR-brane (i.e. with a proper width $\ll \rho$) would change nothing to the qualitative discussion on the warp transmutation of scales which is discussed in the following. The goal of this spinless bulk field is just to model some generic 5D field coupled to a brane-localized scalar (the spin degrees of freedom of this bulk field are irrelevant for the following discussion).

\paragraph{Warp Transmutation of Scales:}
By using the same method as in Section~\ref{KK_flat_toy}, the study of the KK dimensional reduction to 4D of the free local field $\Phi(x, y)$ can be found in Ref.~\cite{Goldberger:1999wh} in the case of a RS throat. The KK decomposition of the smeared field $\widetilde{\Phi}(x, y)$ is formally the same as in Eq.~\eqref{KK_smeared}, where the smeared KK-mode interaction is
\begin{equation}
\widetilde{\phi}_n (x) = e^{\eta^{\prime 2} \partial_\mu^2} \, \phi_n(x) \, ,
\end{equation}
where $\eta^\prime = 1 / \Lambda_\eta^\prime$, with $\Lambda_\eta^\prime = e^{-A(\pi \rho)} \, \Lambda_\eta$. One needs also to rescale the brane field $H(x)$, in order to get a canonically normalized kinetic term \cite{Randall:1999ee}: $h(x) = e^{-A(\pi \rho)} \, H(x)$, such that the 4D action involving only the field $h(x)$ is
\begin{equation}
S_h = \int d^4x \left[ - \dfrac{1}{2} \, h \left( \partial_\mu^2 - \mu_h^2 \right) h + \dfrac{\lambda}{4!} \, \widetilde{h}^4 \right] \, ,
\end{equation}
where $\mu_h = e^{-A(\pi \rho)} \, \mu_H$. For a warp factor satisfying $A(\pi \rho) \gg 1$, both $\mu_H^2$ and $\Lambda_\eta$ appear redshifted from their natural 5D values: this result generalizes the warp transmutation of the RS1 model to the case of the WNL scale\footnote{For the warped throats in string theory, where $\ell_s \sim \eta$, the mass threshold of the Regge excitations is also redshifted \cite{Reece:2010xj}, which means that the string scale $M_s$ is indeed warped down, and it confirms the consistency of the results of this article from a top-down approach.}. It is also useful to introduce the redshifted nonperturbative scale $\Lambda_{IR} = e^{-A(\pi \rho)} \, \Lambda_{UV}$ on the IR-brane. In the following, the scale hierarchy of the model is studied in the particular case of a RS throat.

\paragraph{Randall-Sundrum Throat:}
It is instructive to discuss the particular case of an Euclidean RS throat\footnote{This example is just to illustrate the proposed mechanism with a well-known warp factor, but it is not crucial for the claims of this article to have a RS throat.} (i.e. a slice of EAdS$_5$), where $A(y) = k y$ with the curvature scale $k < \Lambda_{UV}$. The KK-scale is thus $M_{KK} = e^{-A(\pi \rho)} \, k $. The class of WNL theories of gravity considered here are the ones where one gets the usual 5D Einstein-Hilbert gravity of the RS1 model for energies $E \ll \Lambda_{UV} \sim \Lambda_P$. In the following application, only the case $k/\Lambda_\eta \leq 1$ will be considered, such that $\Lambda_P/\Lambda_{UV}$ should still be of the same order of magnitude as in the RS1 model. As discussed in Ref.~\cite{ArkaniHamed:2000ds} in the local RS1 model, the couplings between the 4D fields localized on the IR-brane and the KK-modes become nonperturbative at the scale $\Lambda_{IR}$, which is interpreted as the UV cutoff of the EFT on the IR-brane.

\paragraph{Scale Hierarchy:}
In order to further discuss the scale hierarchy in this WNL RS1 model, one can introduce the following quantities:
\begin{equation}
\kappa_k = \dfrac{k}{\Lambda_{UV}} = \dfrac{M_{KK}}{\Lambda_{IR}} \, , \ \ \ 
\kappa_\mu = \dfrac{\left| \mu_H \right|}{\Lambda_{UV}} = \dfrac{\left| \mu_h \right|}{\Lambda_{IR}} \, , \ \ \ 
\kappa_\eta = \dfrac{\Lambda_\eta}{\Lambda_{UV}} = \dfrac{\Lambda_\eta^\prime}{\Lambda_{IR}} \, ,
\end{equation}
such that:
\begin{itemize}[label=$\spadesuit$]
\item In order to be in the classical regime for the background metric, one needs $\kappa_k \ll 1$. In practice, a mild hierarchy $\kappa_k \sim 10^{-1}$ or $10^{-2}$ is sufficient.
\item The natural value of $\kappa_\mu$ depends on the sensitivity of $|\mu_H|$ on the higher scales $\eta$ and/or $\Lambda_{UV}$.
\item 3 benchmark points will be considered for $\kappa_\eta$, i.e. $\kappa_\eta \rightarrow \infty$, $\kappa_\eta = 1$, and $\kappa_\eta = \kappa_k$.
\end{itemize}
Note that both $\Lambda_\eta$ and $k$ are stable under radiative corrections (as well as their redshifted quantities). A mild hierarchy with $\Lambda_{UV}$ is then natural from a technical naturalness point of view (no fine-tuning), but not from a Dirac naturalness one (i.e. one may want to explain this mild hierarchy by a UV mechanism).

\subparagraph{Benchmark Limit $\kappa_\eta \rightarrow \infty$:}
By keeping $\Lambda_{UV}$ fixed, it means that the WNL scale $\Lambda_\eta$ decouples, and one gets a local RS-like model, so the radiative corrections $\delta \mu_h^2$ to $\mu_h^2$ scale as \cite{Randall:1999ee, ArkaniHamed:2000ds}
\begin{equation}
\delta \mu_h^2 \sim \Lambda_{IR}^2 \ \ 
\Rightarrow \ \ 
\kappa_\mu \sim 1 \ \text{(naturalness)}.
\end{equation}
For a realistic model of EW symmetry breaking, $H(x)$ is promoted to the Higgs field in the EW theory. When the RS1 model was published in 1999 \cite{Randall:1999ee}, $\Lambda_{IR}$ could be as low as few TeV with $k \pi \rho \simeq 34$. Nowadays, since $M_{KK}$ is constrained by the LHC to be at the TeV scale \cite{Workman:2022ynf}, there is a little hierarchy problem, which is usually worse than in other BSM scenarios based on a new symmetry because $\kappa_k \ll 1 \Rightarrow \Lambda_{IR} \gg 1$ TeV and $|\mu_h| \sim \Lambda_{EW}$. 

\subparagraph{Benchmark Point $\kappa_\eta = 1$:}
The WNL scale $\Lambda_\eta$ is redshifted to $\Lambda_\eta^\prime$ on the IR-brane. Any field localized on the IR-brane will then give contributions to $\delta \mu_h^2$ such that \cite{Biswas:2014yia}
\begin{equation}
\delta \mu_h^2 \sim \Lambda_\eta^{\prime 2} \ \ 
\Rightarrow \ \ 
\delta \mu_h^2 \sim \Lambda_{IR}^2
\Rightarrow \ \ \kappa_\mu \sim 1 \ \text{(naturalness)}.
\end{equation}
The situation is thus the same as for the local RS-like models ($\kappa_\eta \rightarrow \infty$).

\subparagraph{Benchmark Point $\kappa_\eta = \kappa_k$:}
In this case, $\eta \sim k \ll \Lambda_{UV}$ (it may be possible that a UV mechanism sets this mild hierarchy). The radiative corrections $\delta \mu_h^2$ to $\mu_h^2$, from the 4D fields localized on the IR-brane, will give \cite{Biswas:2014yia}
\begin{equation}
\delta \mu_h^2 \sim \Lambda_h^2 \ \ 
\Rightarrow \ \ 
\delta \mu_h^2 \sim M_{KK}^2
\Rightarrow \ \ \kappa_\mu \ll 1 \ \text{(naturalness)}.
\end{equation}
In a realistic model with the Higgs field of the EW theory, even if $\Lambda_{IR} \gg M_H$, the Higgs boson mass is protected above the redshifted WNL scale $\Lambda_\eta^\prime$. With $\kappa_\eta = \kappa_k$, one gets a shadow-warped EDS, which should reduce the little hierarchy problem in RS-like models if $\Lambda_\eta^\prime$ is at the TeV scale, which seems to be roughly the current constraints, at least in the toy 4D models of the literature \cite{Biswas:2014yia, Su:2021qvm}. Nevertheless, a complete realistic model is needed for a quantitative statement of the possible remaining fine-tuning, which should be sensitive to the form of the WNL form factor.

\section{Conclusion}
\label{conclusion}
In this article, new possibilities for model building with EDS's and branes have been presented, based on the amazing features of string-inspired nonlocality. In Section~\ref{nonlocal_braneworld}, a WNL braneworld action has been studied for the first time. In this 5D toy model, it has been shown that WNL smears the interactions of 4D fields localized on a $\delta$-like 3-brane, such that one can talk about a fuzzy brane. It is then possible that 4D fields localized on 2 different fuzzy branes can interact, which is not possible in a local 5D EFT, unless they are identified with the quasilocalized zero modes of some 5D fields \cite{Fichet:2019owx}. Therefore, in a WNL framework, it is possible to have quasilocalized 4D modes which do not come with KK-excitations \cite{Nortier:2020vge}: this is a rather unique feature of WNL braneworlds that distinguishes them from their local cousins.
Moreover, KK-excitations of bulk fields whose masses are above the WNL scale have sizable suppressed couplings with the brane-localized fields. If the WNL scale coincides with the KK-scale, one thus expect to be able to suppress the effect of the KK-excitations on the observables, compared to the local braneworld models (shadow EDS). Nevertheless, a better understanding of the phenomenology of this UV nonlocal regime is needed to make solid conclusions.\\
  
In Section~\ref{smeared_hierarchy}, an application of fuzzy branes to flavor physics is proposed. Inspired by the AS models of split fermions \cite{ArkaniHamed:1999dc}, suppressed effective 4D Yukawa couplings are realized by localizing Weyl fermions on different fuzzy brane with a bulk Higgs, such that one should be able to reproduce the same phenomenology as in Ref.~\cite{Mirabelli:1999ks}. It also offers an accurate EFT framework for intersecting brane models. One can imagine other applications in BSM models where feeble interactions are important, such as in dark matter and hidden sectors.\\

In Section~\ref{warped_transmut}, a model with a warped EDS has been considered. A 4D scalar field is localized on the IR-brane, and it has been shown that both the mass of the scalar field and its WNL scale undergo a warp transmutation à la RS. This feature provides a built-in perturbative UV cutoff to stabilize physical scales provided by a Higgs mechanism, such as in the standard EW theory.

\acknowledgments
I thank Anish Ghoshal, Anupam Mazumdar, Ratul Mahanta, Leonardo Modesto and Nobuchika Okada for useful discussions on WNL field theories.

\appendix

\section{Notations \& Conventions}
\label{conventions}
\begin{itemize}[label=$\spadesuit$]
\item The WNL field theories considered in this article must be defined on spacetime with Euclidean signature $(++ \cdots +)$. Indeed, when one computes $S$-matrix elements, only the momenta of the external states are analytically continued to Minkowskian signature $(-+\cdots+)$ à la Efimov in order to preserve unitarity \cite{Efimov:1965mnl, Efimov:1966ylf, Pius:2016jsl, DeLacroix:2018arq, Carone:2016eyp, Briscese:2018oyx, Chin:2018puw, Briscese:2021mob, Koshelev:2021orf, Buoninfante:2022krn}.
\item The Gaussian function $\delta_\eta^{(d)}(\xi)$ on $\mathbb{R}^d$ with $d \in \mathbb{N}^*$ (noted simply $\delta_\eta(\xi)$ when $d=1$) can be expressed as the kernel\footnote{Usually called the heat kernel.} of an infinite-derivative operator:
\begin{equation}
\forall \xi \in \mathbb{R}^d, \ \delta_\eta^{(d)}\left( \xi \right) 
= e^{\eta^2 \Delta_d} \, \delta^{(d)} \left( \xi \right)
= \left( \dfrac{1}{4 \pi \eta^2} \right)^{d/2} \exp \left( - \left\| \dfrac{\xi}{2 \eta} \right\|^2 \right) \, ,
\label{heat_kernel}
\end{equation}
where $\Delta_d$ is the Laplacian on $\mathbb{R}^d$, $\| \cdot \|$ is the Euclidean norm, and the Dirac generalized function\footnote{Generalized function are also usually called distributions in the literature \cite{Schwartz:1966}.} $\delta^{d}(\xi)$ on $\mathbb{R}^d$ (of mass dimension $d$, and noted simply $\delta(\xi)$ when $d=1$) is normalized as
\begin{equation}
\int d^d \xi \ \delta^{(d)} (\xi) = 1 \, .
\end{equation}
\item In this study, the analysis is limited to spacetimes with 1 flat/warped EDS, whose metric can be put in the form:
\begin{equation}
ds^2 = g_{MN}(y) \, dx^M dx^N = g_{\mu \nu}(y) \, dx^\mu dx^\nu + dy^2 \, , \ \ \ 
g_{\mu \nu}(y) = e^{-2 A(y)} \, \delta_{\mu \nu} \, ,
\label{warped_metric}
\end{equation}
where $\mu \in \llbracket 0, 3 \rrbracket$, $M \in \llbracket 0, 4 \rrbracket$, and $e^{-A(y)}$ is the warp factor. One has $A(y) = 0$ in the flat case, and $A(y)$ is an increasing function in the warped case. In this choice of coordinate system, both the determinant of the 5D metric, and the one of the induced metric on a 3-brane at the coordinate $y$, are given by $g=e^{-8 A(y)}$. The 5D Laplacian can thus be split as
\begin{equation}
\Delta = \Delta_\parallel + \Delta_\perp \, , 
\end{equation}
such that when it acts on a scalar field $\Phi(x, y)$, one has
\begin{equation}
\Delta_\parallel \Phi = g^{\mu \nu}(y) \, \partial_\mu \partial_\nu \Phi
= e^{2 A(y)} \partial_\mu^2 \Phi \, ,
\end{equation}
and
\begin{equation}
\Delta_\perp \Phi = \sqrt{\dfrac{1}{g}} \, \partial_y \left( \sqrt{g} \, \partial_y \Phi \right)
= e^{4A(y)} \, \partial_y \left( e^{-4A(y)} \, \partial_y \Phi \right) \, ,
\end{equation}
where $\partial_\mu^2 = \partial_\mu \partial_\mu$ is the Laplacian on $\mathbb{R}^4$.
\item The orbifold $S^1 / \mathbb{Z}_2$ is obtained by modding out the circle $S^1$ of radius $\rho \in \mathbb{R}^*_+$ by the group $\mathbb{Z}_2$, cf. Ref.~\cite{Raychaudhuri:2016}. A point on $S^1$ is labeled by the coordinate $y \in (-\pi \rho, \pi \rho]$. There are 2 fixed points localized at $y = 0, \pi \rho$, which have the features of $\delta$-like 3-branes. This orbifold is topologically equivalent to the interval $[0, \pi \rho]$. The Dirac generalized function $\delta(y)$ (mass dimension 1) on the circle $S^1$ is normalized as
\begin{equation}
\oint dy \ \delta(y) = 1 \, .
\end{equation}
\item The 4D Euclidean Dirac matrices are taken in the Weyl representation:
\begin{equation}
\gamma_\mu =
\begin{pmatrix}
0 & -i \sigma_\mu^- \\
i \sigma_\mu^+ & 0
\end{pmatrix},
\label{gamma_1}
\end{equation}
with
\begin{equation}
\sigma_\mu^\pm = \left( \mp i, \sigma_i \right) \, , \ \ \ 
\left\{ \gamma_\mu, \gamma_\nu \right\} = 2 \delta_{\mu\nu} \, ,
\end{equation}
where $\left( \sigma_i \right)_{i \in\llbracket 1, 3 \rrbracket}$ are the 3 Pauli matrices:
\begin{equation}
\sigma_1 =
\begin{pmatrix}
0 & 1 \\
1 & 0
\end{pmatrix},
\phantom{000}
\sigma_2 =
\begin{pmatrix}
0 & -i \\
i & 0
\end{pmatrix},
\phantom{000}
\sigma_3 =
\begin{pmatrix}
1 & 0 \\
0 & -1
\end{pmatrix}.
\end{equation}
A 4D Euclidean Dirac spinor $\psi$ can be decomposed into its chiral components:
\begin{equation}
\psi =
\begin{pmatrix}
\psi_L \\
\psi_R
\end{pmatrix} \, .
\end{equation}
The operator $\cancel{\partial}$ is defined as
\begin{equation}
\cancel{\partial} \psi (x) = \gamma_\mu \partial_\mu \psi (x) \, , \ \ \ 
\cancel{\partial} \psi_L (x) = \sigma_\mu^+ \partial_\mu \psi_L(x) \, , \ \ \ 
\cancel{\partial} \psi_R (x) = \sigma_\mu^- \partial_\mu \psi_R(x) \, .
\label{cancel_partial}
\end{equation}
\end{itemize}

\bibliographystyle{biblistyle}

\providecommand{\href}[2]{#2}\begingroup\raggedright\endgroup

\end{document}